\documentclass[aps,prx,twocolumn,superscriptaddress,showpacs]{revtex4}

\usepackage{eurosym}
\usepackage{amsfonts}
\usepackage{amssymb}
\usepackage{amsmath}
\usepackage{graphicx}
\usepackage{color}

\begin{document}

\title{Manipulating charge-density-wave in $1T$-TaS$_{2}$ by charge carrier doping: A first-principles investigation}

\author{D. F. Shao}
\email[The authors contributed equally to this work.]{}

\affiliation{Key Laboratory of Materials Physics, Institute of Solid State Physics,
Chinese Academy of Sciences, Hefei 230031, People's Republic of China}

\author{R. C. Xiao}
\email[The authors contributed equally to this work.]{}

\affiliation{Key Laboratory of Materials Physics, Institute of Solid State Physics,
Chinese Academy of Sciences, Hefei 230031, People's Republic of China}

\affiliation{University of Science and Technology of China, Hefei, 230026, People's Republic of China}

\author{W. J. Lu}

\email{wjlu@issp.ac.cn}
\affiliation{Key Laboratory of Materials Physics, Institute of Solid State Physics,
Chinese Academy of Sciences, Hefei 230031, People's Republic of China}

\author{H. Y. Lv}
\affiliation{Key Laboratory of Materials Physics, Institute of Solid State Physics,
Chinese Academy of Sciences, Hefei 230031, People's Republic of China}

\author{J. Y. Li}
\affiliation{Key Laboratory of Materials Physics, Institute of Solid State Physics,
Chinese Academy of Sciences, Hefei 230031, People's Republic of China}

\affiliation{University of Science and Technology of China, Hefei, 230026, People's Republic of China}

\author{X. B. Zhu}
\affiliation{Key Laboratory of Materials Physics, Institute of Solid State Physics,
Chinese Academy of Sciences, Hefei 230031, People's Republic of China}

\author{Y. P. Sun}

\email{ypsun@issp.ac.cn}

\affiliation{High Magnetic Field Laboratory, Chinese Academy of Sciences, Hefei
230031, People's Republic of China}

\affiliation{Key Laboratory of Materials Physics, Institute of Solid State Physics,
Chinese Academy of Sciences, Hefei 230031, People's Republic of China}

\affiliation{Collaborative Innovation Center of Microstructures, Nanjing University, Nanjing 210093, China }

\begin{abstract}
The transition metal dichalcogenide (TMD) $1T$-TaS$_{2}$ exhibits a rich set of charge density wave (CDW) orders. Recent investigations suggested that using light or electric field can manipulate the commensurate (C) CDW ground state. Such manipulations are considered to be determined by the charge carrier doping. Here we simulate by first-principles calculations the carrier doping effect on CCDW in $1T$-TaS$_{2}$. We investigate the charge doping effects on the electronic structures and phonon instabilities of $1T$ structure and analyze the doping induced energy and distortion ratio variations in CCDW structure. We found that both in bulk and monolayer $1T$-TaS$_{2}$,  CCDW is stable upon electron doping, while hole doping can significantly suppress the CCDW, implying different mechanisms of such reported manipulations. Light or positive perpendicular electric field induced hole doping increases the energy of CCDW, so that the system transforms to NCCDW or similar metastable state. On the other hand, even the CCDW distortion is more stable upon in-plain electric field induced electron injection, some accompanied effects  can drive the system to cross over the energy barrier from CCDW to nearly commensurate (NC) CDW or similar metastable state. We also estimate that hole doping can introduce  potential superconductivity with $T_{c}$ of $6\sim7$ K. Controllable switching of different states such as CCDW/Mott insulating state, metallic state, and even the superconducting state can be realized in $1T$-TaS$_{2}$, which makes the novel material have very promising applications in the future electronic devices. 

\end{abstract}

\maketitle

\section{Introduction}

Materials with correlated electrons exhibit some of the most intriguing quantum states in condensed matter physics \cite{Imada-RMP,Dagotto-RMP,Lee-RMP}. Since the number of electric charge carriers essentially determines such states, external electric or light field can be applied for controllable manipulations. Stability of electric field or light induced states has been demonstrated in some novel systems, where switching occurs between neighboring equilibrium thermodynamic states \cite{Ohno-Nature,Yamada-Science,Gu-Science,Glover-PRL,Takubo-PRL,Zakery}. This powerful characteristic can be applied in electric devices such as transistors and memories, which are of great importance to not only the fundamental physics research but also the information processing technology \cite{Nakano-Nature}.   Correlated materials with a rich set of quantum states delicately balanced on a similar energy scale will be promising platforms to realize such devices.

\begin{figure}
\includegraphics[width=0.99\columnwidth]{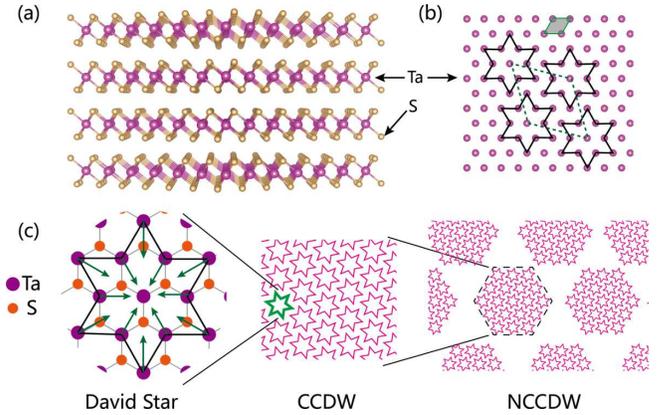}\caption{\label{Fig_1_structure}(a) Crystal structure of $1T$-TaS$_{2}$. (b) Top view of Ta-Ta plane of $1T$-TaS$_{2}$. The unit cells of conventional $1T$ structure and low temperature CCDW structure are denoted with green solid and dashed lines, respectively. The black hexagrams are the so-called ``David Star'' pattern in CCDW phase. (c) Schematic diagrams of ``David Star'' (left), CCDW (middle), and NCCDW (right) in $1T$-TaS$_{2}$.}
\end{figure}

The transition metal dichalcogenide (TMD) $1T$-TaS$_{2}$ is one of the promising candidate due to the multiple competing ground states in it \cite{Wilson-CDW-review}. As shown in Fig. \ref{Fig_1_structure} (a), $1T$-TaS$_{2}$ shows a CdI$_{2}$-type layered crystal structure with Ta atoms octahedrally coordinated by S atoms. A unit layer consists of one Ta layer sandwiched between two S layers. At low temperatures, strong $\textit{\textbf{q}}$-dependent electron-phonon coupling induced periodic lattice distortion makes a $\sqrt{13}\times\sqrt{13}$ superlattice \cite{AmyLiu-1T-TaS2,AmyLiu-1T-TaSe2,TaSeTe-Arxiv}, in which Ta atoms displace to make ``David Star'' clusters (Fig. \ref{Fig_1_structure} (b)). The outer twelve atoms within each star move slightly towards the atom at the center, leading to the commensurate  charge-density-wave (CCDW) ground state. In particular, in CCDW state, the correlation effect of $5d$ electrons of Ta atoms turns the system into Mott insulating state \cite{Wilson-CDW-review,Darancet-PRB,Ang-PRL,Ang-PRB}. Upon heating to 225 K, it undergoes a sequence of first order phase transition to a nearly commensurate (NC) CDW. The NCCDW phase is composed of metallic incommensurate (IC) network and Mott insulating CCDW domains. The CCDW domains shrink upon heating and finally disappear at 355 K, while the system transforms to ICCDW state. The standard metallic $1T$ structure appears above 535 K. Moreover, when CCDW state is suppressed, superconductivity emerges in this system \cite{Sipos-1T-TaS2-pressure,LLJ-EPL,LY-APL}. One can expect that the controllable switching between those states will be helpful for figuring out the mechanism of CDW and superconductivity, and realizing the high performance memory and transistor in future technology. To meet this goal, many groups performed investigations on this novel material. Yu et al. reported a gate-controlled Li ion intercalation can suppress CCDW and introduce superconducivity \cite{Yu-Li-intercalation}. Tsen et al. \cite{Tsen}, Hollander et al. \cite{Hollander}, Yoshida et al. \cite{Yoshida-Sience-Advance},  and Mihailovic et al. \cite{Vaskivskyi-arxiv,Vaskivskyi-Science-advance} reported the in-plane electric field induced transition from CCDW state to NCCDW or some metastable hidden state. Cho et al. applied perpendicular electric pulse on $1T$-TaS$_{2}$ and found positive electric pulse can introduce IC network to suppress Mott state at low temperature \cite{Cho-STM}. Moreover, Zhang et al. \cite{ZGSheng}, Mihailovic et al. \cite{Stojchevska,Vaskivskyi-Science-advance}, and Han et al. \cite{Han-science-advance}  suggested that light can introduce a transition from CCDW state to NCCDW or hidden state as well. In all those manipulations, the transitions are considered to be determined by the charge carrier doping. However, the mechanisms of such transitions are not fully clear yet.

In this work, we simulated by first-principles calculations the charge carrier doping effect on CDW in $1T$-TaS$_{2}$. We found that CCDW is stable upon  electron doping, while hole doping significantly suppresses CCDW instability, implying different mechanisms of recently reported electric  and photoelectric manipulations of CDW in $1T$-TaS$_{2}$. We figured out such mechanism by analysis of carrier doping effects. Furthermore, we show that superconductivity with $T_{c}$ about $6\sim7$ K can be introduced by hole doping in the system.

\section{Methods}

The density functional theory (DFT) calculations were carried out using QUANTUM ESPRESSO package \cite{QE} with ultrasoft pseudopotentials. The exchange-correlation interaction was treated with the generalized gradient approximation (GGA) with PW91 parametrization \cite{PW91}. The energy cutoff for the plane-wave basis set was 35 Ry. The Marzari-Vanderbilt Fermi smearing method \cite{MV} with a smearing parameter of $\sigma=0.02$ Ry was used for the calculations of the total energy and electron charge density. To simulate the monolayer, a vacuum layer more than 15 $\mathrm{\AA}$ was introduced. For the bulk sample, brillouin zone sampling is performed on the Monkhorst-Pack (MP) mesh \cite{MP} of $16\times16\times8$, while a denser $32\times32\times16$ grid was used in the electron phonon coupling calculations. Phonon dispersions were calculated using density functional perturbation theory (DFPT) \cite{DFPT} with an $8\times8\times4$ mesh of $q$-points. For the monolayer sample, $k$-points grids of  $64\times64\times1$ and $32\times32\times1$ and $q$-points grids of $8\times8\times1$ are used. Since the frequency and stability of soft phonon modes were found to be very sensitive to the plane-wave cutoff for the charge density, a large cutoff of 1500 Ry was used for the charge density in the phonon calculations \cite{AmyLiu-1T-TaS2}.  All the calculation parameters were well tested. In the investigation of carrier doping effects, the electron/hole doping was simulated  by increasing/decreasing the total electron numbers of the system,   together with a compensating uniform positive/negative background to maintain the charge neutrality. The crystal structure of each doped sample was optimized with respect to lattice parameters and atomic positions. For comparisons, we also calculated some doped samples by using the lattice parameters of pristine TaS$_{2}$ and relaxing the atomic positions, as described later.

\section{Results and discussions}

\begin{figure*}
	\includegraphics[width=0.6\textwidth]{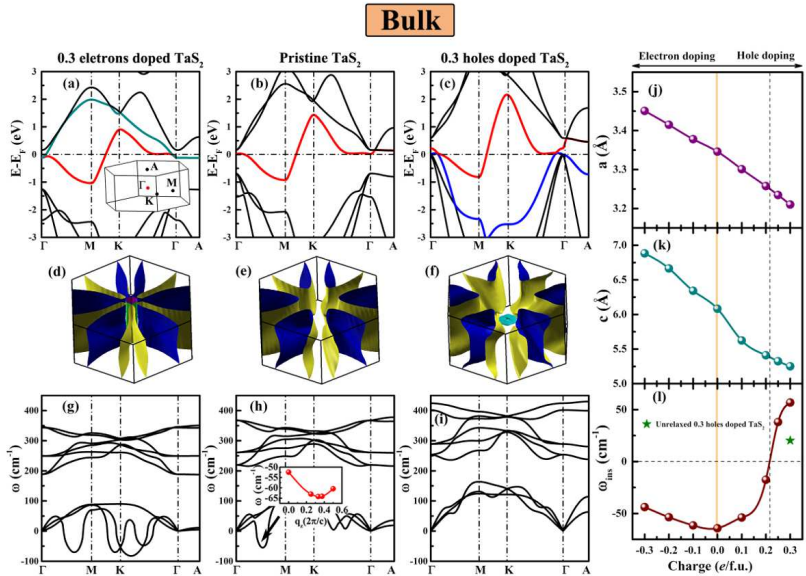}
	
	\caption{\label{Fig_2_bulk} The doping effects in bulk TaS$_{2}$ investigated using the high symmetry $1T$ structure. For the doped TaS$_{2}$ with  $n=0.3$ electrons/f.u., the pristine TaS$_{2}$, and the doped TaS$_{2}$ with   $n=0.3$ holes/f.u., (a), (b), and (c) are the band dispersions of those samples, (d), (e), and (f) are the Fermi surfaces, and (g), (h), and (i) are the phonon dispersions. The high-symmetry points are denoted in the inset of (a). The inset of (h) shows the $q_{z}$ dependence of unstable acoustic branch in pristine TaS$_{2}$. (j) and (k) show the charge carrier doping induced variations of lattice parameters $a$ and $c$, respectively. (l) shows the phonon frequency variations of mode near CCDW instability. The green star in (l) denotes the phonon frequency of mode near CCDW instability of the doped bulk $1T$-TaS$_{2}$ with $n=0.3$ holes/f.u. calculated using the lattice parameters of the prinstine TaS$_{2}$. The negative charge means the electron doping, while the positive charge means the hole doping.}

\end{figure*}

The low symmetry CDW structure is usually considered as the high symmetry phase with distortion introduced by some instability. Therefore, we firstly investigated carrier doping effects in the bulk $1T$-TaS$_{2}$. The electronic structures and phonon properties for the samples with different doping level were calculated.  As examples, Figure \ref{Fig_2_bulk} shows the calculated band structures and Fermi surfaces of the pristine bulk $1T$-TaS$_{2}$, the doped bulk $1T$-TaS$_{2}$ with the doping level of  $n=0.3$ electrons/f.u., and  the doped bulk $1T$-TaS$_{2}$ with $n=0.3$ holes/f.u. As expected, electron doping increases the lattice parameters and hole doping shrinks the lattice (Figs. \ref{Fig_2_bulk} (j) and (k)). For the pristine TaS$_{2}$, our results are of good agreement with the previous calculations \cite{Wilson-CDW-review}. There is a gap of $\sim0.7$ eV below Fermi energy ($E_{F}$) (Fig. \ref{Fig_2_bulk} (b)). In the $\Gamma$-$A$ direction, the two bands around the gap are nearly flat due to the quasi-2D nature of the layered structure. The band above the gap crosses Fermi energy ($E_{F}$), forming a 2D electron pocket around $M$ point (Figs. \ref{Fig_2_bulk} (b) and (e)). Such gap increases upon electron doping and decreases upon hole doping (Figs. \ref{Fig_2_bulk} (a) and (c)). Electron doping increases $E_{F}$. As shown in Figs. \ref{Fig_2_bulk} (a) and (d), for the doped bulk $1T$-TaS$_{2}$ with $n=0.3$ electrons/f.u. the rise of $E_{F}$ expands and opens up the original electron pockets around $M$, leaving hole pockets centered at $K$ point. Besides the original band crossing $E_{F}$, a band with higher energy starts to cross $E_{F}$, forming a 2D cylinder-like electron pocket around the zone center.  On the contrary, the hole doping enhances the dispersion in the $\Gamma$-$A$ direction and weakens the quasi-2D nature by shrinking the lattice and decreasing the interlayer distance.  As shown in Figs. \ref{Fig_2_bulk} (c) and (f),   for the doped bulk $1T$-TaS$_{2}$ with $n=0.3$ holes/f.u., the bands close to $E_{F}$ are not flat anymore. Hole doping reduces $E_{F}$.  The decrease of $E_{F}$ shrinks the original electron pockets around M point. Moreover, a lower energy band starts to cross $E_{F}$, forming a 3D hole pocket around $\Gamma$ point. 

The phonon instability of the high symmetry structure is considered to be directly related to the CDW distortion: At high temperatures, the phonon of the high symmetry structure softens at  CDW vector ($\textit{\textbf{q}}_{CDW}$).  Above the transition temperature the phonon frequency near $\textit{\textbf{q}}_{CDW}$ drops but does not go to zero. Just below the transition temperature the phonon frequency near $\textit{\textbf{q}}_{CDW}$ is imaginary, meaning there is a restructuring of  lattice with a superlattice vector of $\textit{\textbf{q}}_{CDW}$ \cite{PNAS-CDW-classfication}. Therefore,  phonon dispersion without imaginary frequency implies that the structure is stable compared to CDW structure.  The phonon calculation is proved to be an effective method to simulate the CDW instability: The calculated phonon dispersions show instability just locating at  $\textit{\textbf{q}}_{CDW}$ of some TMDs \cite{AmyLiu-1T-TaS2,AmyLiu-1T-TaSe2,TaSeTe-Arxiv,Johannes-NbSe2,Battaglia-NbTe2,Bianco-TiSe2}. More specially, for the present $1T$-TaS$_{2}$, experimental report showed CDW can be suppressed by pressure \cite{Sipos-1T-TaS2-pressure}, which is correctly simulated by Liu's phonon calculation \cite{AmyLiu-1T-TaS2}. In this work we also performed the phonon calculations on each sample. For the pristine $1T$-TaS$_{2}$, our calculation is of good agreement with Liu's calculation (Fig. \ref{Fig_2_bulk} (h)). The phonon dispersion show instability very close to the CCDW vector $\textit{\textbf{q}}_{CCDW}=\frac{3}{13}\textit{\textbf{a}}^{\textbf{*}}+\frac{1}{13}\textit{\textbf{b}}^{\textbf{*}}$. This instability persists at all values of $\textit{\textbf{q}}_{z}$, as shown in inset of Fig. \ref{Fig_2_bulk} (h). For the electron doped sample, the acoustic branches become more unstable. As shown in Fig. \ref{Fig_2_bulk} (g), the unstable modes in the doped bulk $1T$-TaS$_{2}$ with with $n=0.3$ electrons/f.u. appear in $K$-$M$ and $K$-$\Gamma$ directions, indicating the area of instability are largely expanded. On the contrary, the hole doping significantly stabilizes lattice. As shown in Fig. \ref{Fig_2_bulk} (i), no unstable mode can be found in the phonon dispersion of the doped bulk $1T$-TaS$_{2}$ with $n=0.3$ holes/f.u. Since the lowest mode in the pristine TaS$_{2}$ locates near $\textit{\textbf{q}}=\frac{3}{13}\textit{\textbf{a}}^{\textbf{*}}+\frac{1}{13}\textit{\textbf{b}}^{\textbf{*}}+\frac{1}{3}\textit{\textbf{c}}^{\textbf{*}}$, we used such mode as an indicator of doping effect on CCDW in TaS$_{2}$. The frequency variation of such mode under doping is shown in Fig. \ref{Fig_2_bulk} (l). One can note that upon electron doping the mode is always unstable in bulk $1T$-TaS$_{2}$, while hole doping significantly suppresses the instability. According to our calculation, the lattice becomes completely stable when the doping level is higher than $n=0.2$ electrons/f.u.   

Since hole doping shrinks the lattice as the high pressure does, we should figure out the suppression of CCDW under hole doping is majorly attributed to whether the lattice compression or to the pure doping effect. We calculated a hypothetic  doped $1T$-TaS$_{2}$ with n=0.3 holes/f.u., in which the lattice parameters are fixed to those of the undoped pristine $1T$-TaS$_{2}$. As shown in Fig. \ref{Fig_2_bulk} (l), CCDW can be suppressed by just doping holes into $1T$-TaS$_{2}$ without changing its lattice volume. That demonstrates the suppression of CCDW in the present case is predominately by the hole doping effect.

\begin{figure}
	\includegraphics[width=0.99\columnwidth]{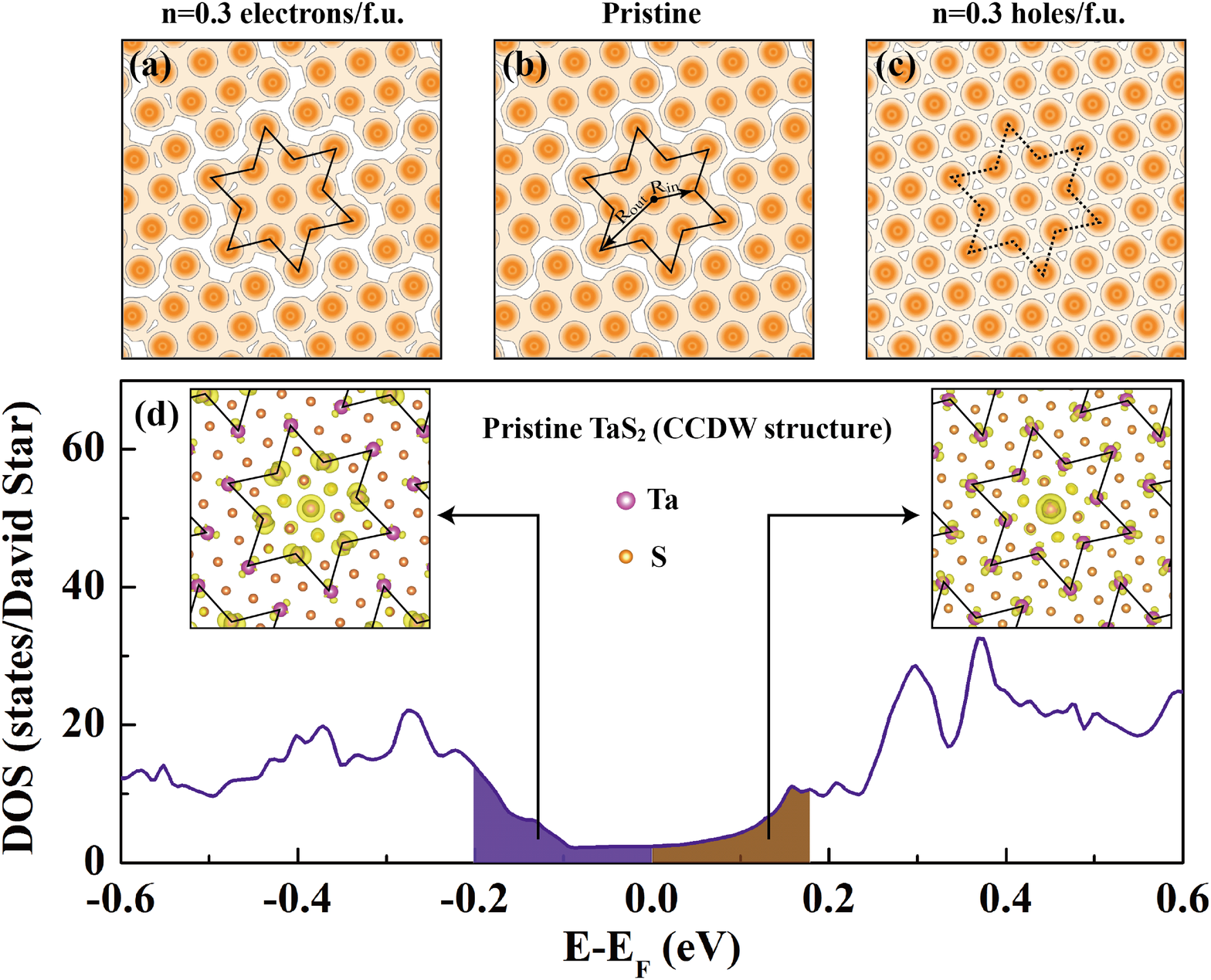}
	
	\caption{\label{Fig_3_charge}The doping effects in the bulk TaS$_{2}$ investigated using the low symmetry CCDW structure. (a), (b), and (c) are the charge density in Ta-Ta plane of the doped TaS$_{2}$ with $n=0.3$ electrons/f.u., pristine TaS$_{2}$, and the doped TaS$_{2}$ with $n=0.3$ holes/f.u., respectively. (d)  DOS of  the pristine TaS$_{2}$ directly calculated in CCDW structure. The colored areas below and above $E_{F}$ denote the electron density which can be integrated to one electron. The related electron charge densities are plotted in the insets of (d).}
\end{figure}

Furthermore, we investigated the doping induced energy and the distortion variations in the CCDW state  
for the pristine TaS$_{2}$, the doped TaS$_{2}$ with $n=0.3$ electrons/f.u., and the doped TaS$_{2}$ with $n=0.3$ holes/f.u. We defined the CCDW formation energy $\Delta E$ as
\begin{equation}
\Delta E=E_{CCDW}-E_{1T},
\end{equation}
where $E_{CCDW}$ and $E_{1T}$ are the total energies of the  relaxed CCDW structure and the $1T$ structure.  The distortion rations $dr$ can be expressed as

\begin{equation}
dr_{in}=\frac{a-r_{in}}{a}\times100\%,
\end{equation}
and 
\begin{equation}
dr_{in}=\frac{\sqrt{3}a-r_{out}}{\sqrt{3}a}\times100\%,
\end{equation} where $a$ is the in-plane lattice parameter of the undistorted $1T$ structure, $r_{in}$ and  $r_{in}$ are radius of the inner and outer circles of ``David Star'', as shown in Fig. 3 (b). For $1T$-TaS$_{2}$ with CCDW structure,  the layers stacking order is not clear yet. Therefore, we simply construct the $\sqrt{13}\times\sqrt{13}\times1$ supercell with 39 atoms to simulate the CCDW structure. We relaxed the atomic positions using the lattice parameters   $\sqrt{13}a$ and $c$, where $a$ and $c$ are from Figs. \ref{Fig_2_bulk} (j) and (k), respectively. In the relaxed CCDW structures, we found that for the  pristine TaS$_{2}$ and  the doped TaS$_{2}$ with  $n=0.3$  electrons/f.u. (Figs. \ref{Fig_3_charge} (b) and (a)), the distortion ratios are large, while the formation energies of those are negative. On the other hand, the distortion ratios of  the doped TaS$_{2}$ with  $n=0.3$ holes/f.u. are nearly zero (Fig. \ref{Fig_3_charge} (c)), and the energy difference between the relaxed supercell and the $1T$ unit cell is very small ($<1$ meV), i.e. the CCDW structure relaxes to high symmetry $1T$ structure.  We also relaxed the atomic positions with $a$ and $c$  fixed to those in the pristine TaS$_2$ to prevent the volume variation. With the lattice parameters of the pristine TaS$_2$, the doped TaS$_2$ with $n=0.3$ electrons/f.u. can be seen as under pressure. However, the CCDW is still stable. On the other hand, without doping introduced volume shrinking, in the  doped TaS$_2$ with $n=0.3$ holes/f.u. the CCDW is still not  favorable in energy. Such results of the CCDW supercell coincide well with our deductions based on the analysis of the phonon instabilities of the $1T$ unit cell.

If we simply consider that the roles of electron/hole doping on the electronic structures are to increase/decrease $E_{F}$, we can estimate  the population of the added/removed electrons when electron/hole doping and understand the calculated results more clearly. Figure \ref{Fig_3_charge} (d) shows the density of states (DOS) near $E_{F}$ of the pristine TaS$_2$ in the CCDW structure  calculated directly by DFT. Based on the integrations of the DOS upwards/downwards from $E_{F}$, we can see that if one electron is doped into a ``David Star'' (the doping level is equal to $n= 1/13$ electron/f.u.), it will be mainly added into the center of the ``David Star''. Obviously, it will enhance the clustering of the charge density. Therefore, CCDW is stable upon electron doping. On the contrary, doping one hole into a ``David Star'' will notably decrease the charge density at the center and inner atoms of the  ``David Star'', which will weaken the charge density clustering. Therefore, CCDW is suppressed upon hole doping.

\begin{table*}

	\caption{\label{CCDW}  CCDW formation energy $\Delta E$ and  distortion ratios	$dr_{in}$ and $dr_{out}$ in the pristine TaS$_{2}$, the doped TaS$_{2}$ with  $n=0.3$ electrons/f.u., and the doped TaS$_{2}$ with  $n=0.3$ holes/f.u. in CCDW structure. The data after ``$/$'' are calculated by fixing the lattice parameters to those of the prinstine TaS$_{2}$ and relaxing the atomic positions.}

\begin{tabular}{ccccccc}
	\hline 
	&~& $\Delta E$ (meV/f.u.) &~& $dr_{in}$ (\%) &~& $dr_{out}$ (\%)\tabularnewline
	\hline 
	Prinstine  &~& $-13.54$ &~& $5.39$ &~& $3.70$\tabularnewline
	$n=0.3$ electrons/f.u. &~& $-19.37/-6.31$ &~& $6.64/4.77$ &~& $3.78/1.85$\tabularnewline
	$n=0.3$ holes/f.u. &~& $-0.22/0.41$ &~& $0.00/0.00$ &~& $0.00/0.00$\tabularnewline
	\hline 
\end{tabular}
	
\end{table*}

\begin{figure}
\includegraphics[width=0.99\columnwidth]{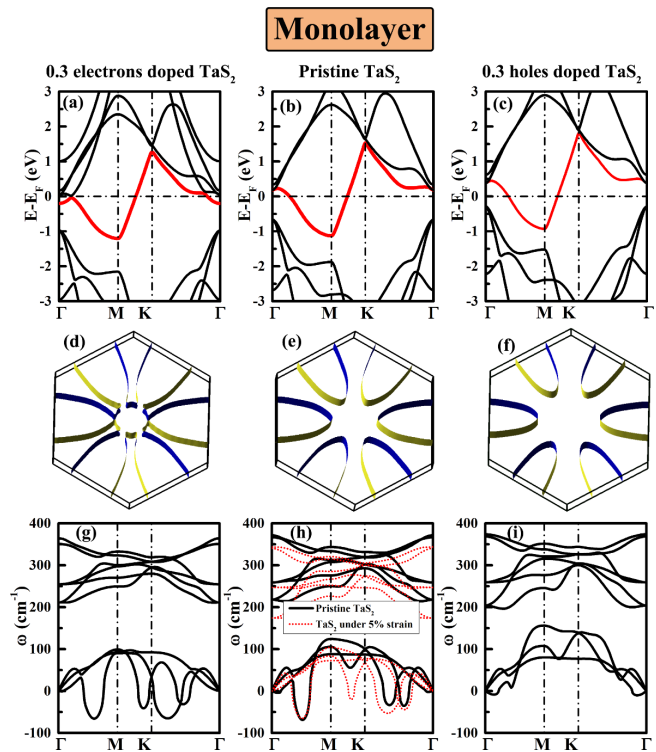}

\caption{\label{Fig_4_monolayer} The doping effects in monolayer $1T$-TaS$_{2}$. For the  doped TaS$_{2}$ with $n=0.3$ electrons/f.u., pristine TaS$_{2}$, and the doped TaS$_{2}$ with $n=0.3$ holes/f.u., (a), (b), and (c) are the band dispersions of those samples, (d), (e), and (f) are the Fermi surfaces, and (g), (h), and (i) are the phonon dispersions, respectively. The red dashed lines show the phonon dispersion of undoped monolayer $1T$-TaS$_{2}$ with tensile strain of 5\%.}
\end{figure}

\begin{figure}
	\includegraphics[width=0.99\columnwidth]{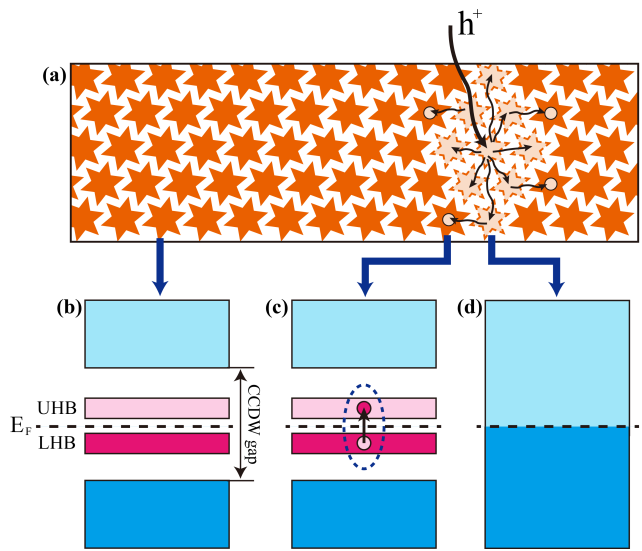}
	
	\caption{\label{Fig_5_hole_mechanism} Schematic picture of suppression of CCDW by hole doping. (a) The situation of TaS$_{2}$  in the CCDW state when holes are doped locally. The solid orange  ``David Stars'' are the area under CCDW distortion. The hollow orange  ``David Stars'' are the area in which only one hole  per a ``David Star'' is doped, which are still under  CCDW distortion.  The dashed light orange ``David Stars'' are the area with more doped holes, in which the CCDW distortion is fully suppressed. The black arrows denote the diffusion of the doped holes. (b) The  schematic band structure of TaS$_{2}$ with CCDW distortion. (c) The  schematic band structure of TaS$_{2}$ with CCDW distortion when one hole per ``David Star'' is doped. (d) The  schematic band structure of TaS$_{2}$ in $1T$ structure.}
\end{figure}

\begin{figure*}
\includegraphics[width=0.7\textwidth]{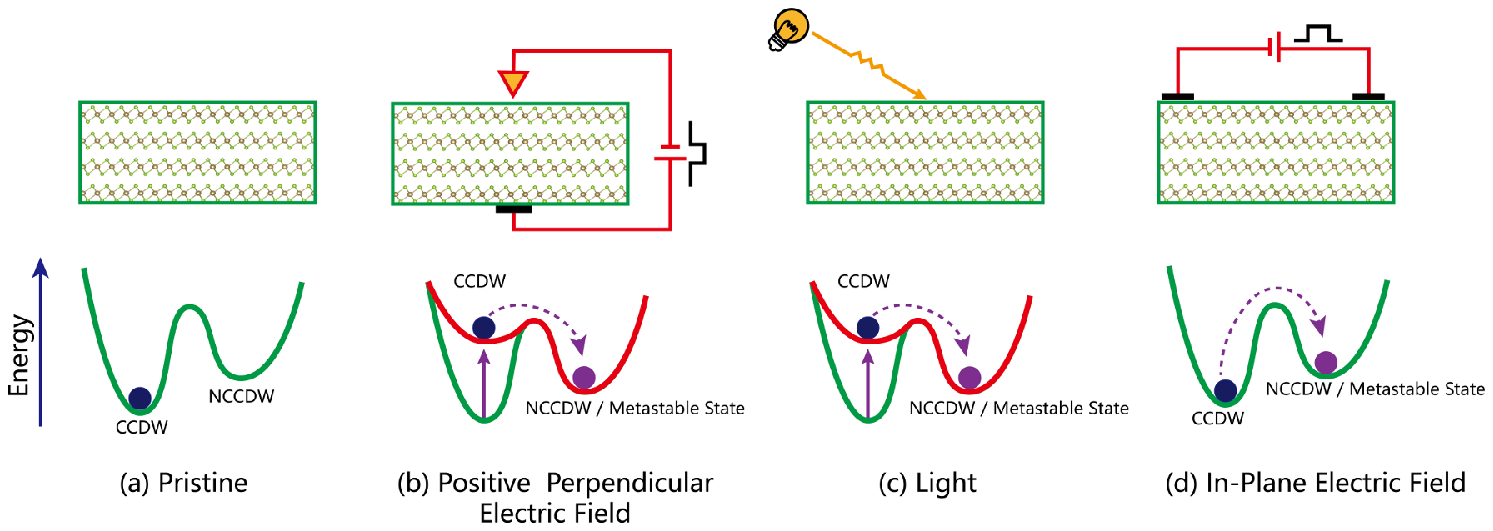}

\caption{\label{Fig_6_mechanism} Schematic diagram of (a) energy in the ground state of $1T$-TaS$_{2}$, mechanism of switching between CCDW and NCCDW/metastable state induced by (b) perpendicular positive electric field, (c) light and (d) in-plane electric field. }
\end{figure*}

We also investigate the doping effect in monolayer $1T$-TaS$_{2}$. As examples,  Fig. \ref{Fig_4_monolayer} shows the calculated electronic structures and phonon properties of the monolayer pristine TaS$_{2}$,  the doped TaS$_{2}$ with $n=0.3$ electrons/f.u., and the doped TaS$_{2}$ with $n=0.3$ holes/f.u in $1T$ structure. For the undoped pristine monolayer $1T$-TaS$_{2}$, the band structure is similar to that of pristine bulk sample. One band crosses $E_{F}$, forming a 2D electron pocket around $M$ point (Figs. \ref{Fig_4_monolayer} (b) and (e)). There is a gap of ~0.8 eV below the band crossing $E_{F}$. In the electrons and holes doped monolayer samples, the gap slightly changes, which is different from the case of bulk samples. For the doped monolayer $1T$-TaS$_{2}$ with $n=0.3$ electrons/f.u., the band structure is slightly different from that of the doped bulk $1T$-TaS$_{2}$ with $n=0.3$ electrons/f.u. (Fig. \ref{Fig_4_monolayer} (a)). Although $E_{F}$ changes with electron doping, there is still only one band crossing $E_{F}$. The 2D cylinder-like electron pocket in the doped bulk $1T$-TaS$_{2}$ with $n=0.3$ electrons/f.u.  does not appear in monolayer sample (Fig. \ref{Fig_4_monolayer} (d)). For the doped monolayer $1T$-TaS$_{2}$ with $n=0.3$ holes/f.u., there is only one band crossing $E_{F}$  (Fig. \ref{Fig_4_monolayer} (c)). Similar to the case of bulk samples, the phonon calculation shows that the CDW instability in the monolayer  $1T$-TaS$_{2}$ cannot be suppressed by electron doping (Fig. \ref{Fig_4_monolayer} (g)), while it can be suppressed under hole doping (Fig. \ref{Fig_4_monolayer} (i)). The result indicates the suppression of CCDW in $1T$-TaS$_{2}$ is not due to the hole doping enhanced band dispersion along $\Gamma$-$A$ direction (Figs. \ref{Fig_2_bulk} (c) and (f)). The suppression should be attributed to the weakening of the electron-phonon coupling at $\textit{\textbf{q}}_{CDW}$ upon hole doping.

Besides, one may note in the monolayer sample, a small instability near $\Gamma$ can be found in the phonon dispersion (Fig. \ref{Fig_4_monolayer} (h)). In a very recent calculation of phonon dispersion of monolayer $1T$-TaS$_{2}$ by Zhang et al. \cite{Zhang-monolayer-TaS2}, there is a similar instability near $\Gamma$. Such instability is consistent
with the instability against long-wavelength transversal waves \cite{G-instability-1,G-instability-2}. This instability is suggest to be fixed by defects, such as ripples or grain boundaries, which do not allow these waves by limiting the size \cite{G-instability-1,G-instability-2,G-instability-3}.  Furthermore, we found tensile strain can suppress such instability near $\Gamma$, which can be easily applied to monolayer materials \cite{Bissett-strain,Park-strain}. As an example, in Fig. \ref{Fig_4_monolayer} (h), the red dashed lines show the phonon dispersion of  $1T$-TaS$_{2}$ under a tensile strain of 5\%, in which the instability near $\Gamma$ is suppressed. The observation indicates that the application of tensile strain is helpful for stabilizing the experimentally exfoliated monolayer or fewlayer $1T$-TaS$_2$.

Based on our calculations of phonon properties in $1T$ structure, CCDW formation energies and distortion ratios in CCDW structure, we can conclude that when the doping level is above $n=0.2$ holes/f.u. (2.6 holes per a ``David Stars''), CCDW can be completely suppressed in $1T$-TaS$_{2}$ for bulk and monolayer, i.e. all the ``David Stars'' should be melted. Experimentally, holes can directly be introduced by a positive electric field perpendicular to the sample \cite{Cho-STM}, or by light \cite{Stojchevska,Vaskivskyi-Science-advance,ZGSheng, Han-science-advance}. One can expect that in reality the doped holes will firstly distribute in a local area and gradually diffuse out, i.e. even the doping level might be much lower than 0.2 holes/f.u., in the local area, the doping level could be high enough to melt the ``David Stars'' in this area and destroy the long-range CCDW, as shown in Fig. \ref{Fig_5_hole_mechanism} (a).  We describe the possible picture of such process here: In $1T$ structure, the electronic structure near $E_{F}$ are formed predominantly from a single Ta $d$ band, which  splits into subbands by the formation of CCDW state. Six of these subbands are fulfilled with 12 electrons per new CCDW unit cell, forming a manifold of occupied states. The 13th leftover electron is localized on the central Ta atom of the ``David Stars'', forming a half-filled subband at $E_{F}$. This half-filled subband further splits by the Coulomb interaction into upper and lower Hubbard bands (UHB and LHB), as shown in Fig. \ref{Fig_5_hole_mechanism} (b). Light or positive perpendicular electric filed can firstly excite one electron from LHB to UHB, and create one hole in LHB (Fig. \ref{Fig_5_hole_mechanism} (c)). In real space, the hole doped by light or positive perpendicular electric filed locates at the center of the  ``David Stars'', leaving a polaron with an excess charge.  The other 12 electrons in this polaron are still star-shaped around the central, thus screening the excess charge. When more holes are doped into the  ``David Stars'', the CCDW distortion is suppressed locally, the structure transforms into $1T$ in the doping area. The splitting subbands merge to a single band again (Fig. \ref{Fig_5_hole_mechanism} (d)).  In this case, the  ``David Stars'' shaped clusters are annihilated and cannot screen the holes any more. So the holes diffuse into neighboring  ``David Stars''. Therefore, upon hole doping in $1T$-TaS$_2$ in CCDW phase, CCDW should firstly transform to a NCCDW or metastable phase composed of CCDW domains and ICCDW network. For example, Cho et al. \cite{Cho-STM} applied a very small positive perpendicular voltage pulse on $1T$-TaS$_{2}$ single crystal sample within a typical scanning tunneling microscope (STM) set-up. Since local hole concentration under the STM tip is largely enhanced, a pulse creates a textured CDW domain of a few tens of nanometers with an irregular domain wall network inside. They considered that such network is consistent with those in the thermally excited NCCDW phase \cite{Cho-STM}. Besides, dI/dV measured  implies the weakening and broadening of the Hubbard states together with the reduction of the Mott gap inside the textured CDW domain induced by positive perpendicular electric filed \cite{Cho-STM}. That is corresponding to our picture: Holes diffuse from the domain wall into the neighboring area inside the domain. Although the concentration of the diffused holes is not high enough to suppress CCDW distortion inside the domain, it can create the hole-electron pair in LHB and UHB to reduce the Mott gap.  Moreover, the photoexcitation of electrons can be considered as directly doping holes into system. Mihailovic's group found a laser pulse can introduce a transition from CCDW to a hidden state in $1T$-TaS$_{2}$ thin flake \cite{Stojchevska,Vaskivskyi-Science-advance}. The Raman spectrum of such hidden state is completely different from that in CCDW and NCCDW phases \cite{Stojchevska}. Such hidden state is demonstrated to be a metastable phase, which is different from NCCDW, but should be composed of CCDW domains and ICCDW network as well \cite{Stojchevska,Vaskivskyi-Science-advance}. Han et al. observed the electron diffraction spots varies from CCDW case to ICCDW case when applying laser to TaS$_{2}$ in CCDW phase \cite{Han-science-advance}. Very recently, Zhang et al. showed laser can introduce a CCDW-NCCDW transition in bulk $1T$-TaS$_{2}$  single crystal \cite{ZGSheng}. On the other hand, in experiments upon hole doping in $1T$-TaS$_{2}$  in high temperature phase, the CCDW transition temperature is significantly lowered  \cite{ZGSheng,Han-science-advance}. Both the two kinds of structural evidences can be well explained by our calculation. For the sake of illustration we draw, we drew a schematic diagram to describe the mechanism of the hole doping induced CCDW-NCCDW/metastable phase transition, as shown in Figs. \ref{Fig_6_mechanism} (a), (b) and (c). In the pristine $1T$-TaS$_{2}$, CCDW phase has the lowest energy (Fig. \ref{Fig_6_mechanism} (a)). The hole doping by perpendicular field or by light can stabilize lattice and  largely increase the energy of CCDW phase. In this case, the system transforms to NCCDW or other metastable phase (Figs. \ref{Fig_6_mechanism} (b) and (c)).

On the other hand, according to our calculation, CCDW lattice distortion is stable upon electron doping. According to the report by Cho et al., negative perpendicular voltage pulses could not make change in CCDW state \cite{Cho-STM}, which is consistent with our estimation. However, some recent experimental works suggest the opposite results. Yu et al. doped electron into $1T$-TaS$_{2}$ by intercalation of Li ions and found CCDW can be suppressed \cite{Yu-Li-intercalation}. The intercalated ions cannot only carry electrons, but also strongly influence the lattice structures and induce disorder, which can suppress CDW in TMDs \cite{Wilson-CDW-review,Cu-TiSe2,Cu-TaS2}. Therefore, one cannot simply attribute the suppression to the electron doping. To demonstrate the pure electron doping effect on CDW, more direct doping experiments by negative perpendicular electric field or liquid-gated method should perform. Some works reported the suppression of CCDW by in-plane electric field \cite{Tsen,Hollander,Vaskivskyi-arxiv,Vaskivskyi-Science-advance, Yoshida-Sience-Advance}. By measuring the temperature dependence of resistivity ($R-T$), Yoshida et al. found the in-plane field cannot affect the NCCDW/ICCDW transition in $R-T$ curve, but can introduce a metastable state with very low resistivity at  low temperatures \cite{Yoshida-Sience-Advance}.  If we consider that the  $R-T$ in CCDW, NCCDW, and ICCDW reflects the related structural characteristics, one can infer that the in-plane field induced the metastable state with different structural characteristics, i.e. the in-plane field can suppress CCDW distortion as well. The suppressions of CCDW by the in-plane field are usually explained as electron injection effect \cite{Hollander,Vaskivskyi-Science-advance}. However, according to our calculation, CCDW could be stable upon electron doping. Therefore, the effect of in-plane field should be more complex. We consider some potential mechanisms of the suppressions by in-plane field: Besides the pure charge carrier injection, one can expect that the in-plane electric field can also force the electrons in the ``David Star'' to be delocalized. Moreover, the in-plane electric field might depinning the CDW, which is suggested in $1T$-TaS$_{2}$ bulk single crystals \cite{depinning}. The suppression of CCDW by in-plane electric field might also  be  due to the thermal activation by local Joule heating as current flows through the material. The real mechanism of the suppression of CCDW by in-plane field needs to be figured out by more experimental and theoretical works. Here we only offer a general description of such process: Although CCDW state might still have the lower energy, the in-plane electric field can drive the system to cross over the energy barrier from CCDW to NCCDW or other metastable phase, as shown in the schematic diagram in Fig. \ref{Fig_6_mechanism} (d).

\begin{figure}
\includegraphics[width=0.99\columnwidth]{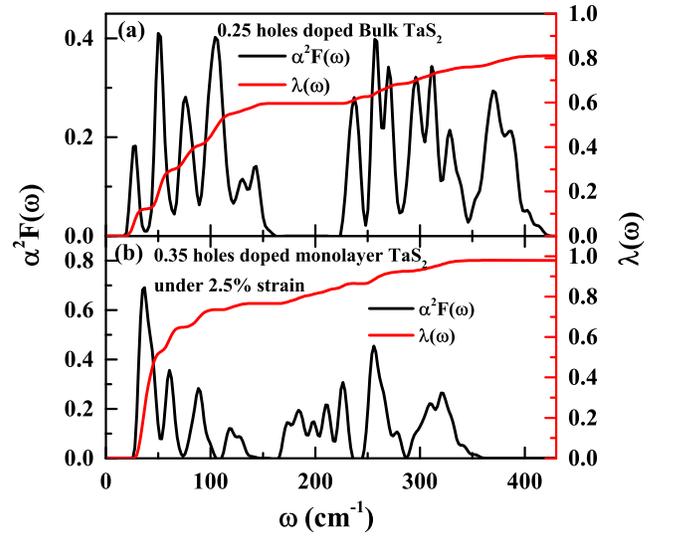}

\caption{\label{Fig_7_a2f} Eliashberg function (left) and the integrated electron-phonon coupling strength (right) for (a) the doped bulk $1T$-TaS$_{2}$ with $n=0.25$ holes/f.u. and (b) the doped monolayer $1T$-TaS$_{2}$ with $n=0.35$ holes/f.u. under a tensile strain of 2.5\%,  respectively. }
\end{figure}

As described above, once long range CCDW is suppressed, the transition between Mott insulating state to metallic state can be observed: When CCDW is suppressed, domain walls show up, which can be seen as conducting channels to induce metallic state \cite{Sipos-1T-TaS2-pressure}. In the CCDW area near the domain walls, the Hubbard states are weakening and broadening and the Mott gap is reduced.  \cite{Cho-STM}. Furthermore, superconductivity can emerge in percolated metallic IC network \cite{Sipos-1T-TaS2-pressure}. While we are not able to directly model the textured IC phase, we can still estimate the superconductivity using $1T$ structure qualitatively. Approximately using  the high symmetry  doped bulk $1T$-TaS$_{2}$ with $n=0.25$ holes/f.u.  and the  doped monolayer $1T$-TaS$_{2}$ with $n=0.35$ holes/f.u. under a tensile strain of 2.5\%, for which all the phonon instabilities are just suppressed, we estimated the potential carrier doping induced superconductivity by the electron-phonon coupling calculation. Figure \ref{Fig_7_a2f} shows the calculated Eliashberg spectral function
\begin{equation}
\alpha^{2}F(\omega)=\frac{1}{N(E_{F})}\underset{\mathbf{k},\mathbf{q},\nu,n,m}{\sum}\delta(\epsilon_{\mathbf{k}}^{n})\delta(\epsilon_{\mathbf{k}+\mathbf{q}}^{m})\mid g_{\mathbf{k},\mathbf{k}+\mathbf{q}}^{\nu,n,m}\mid^{2}\delta(\omega-\omega_{\mathbf{q}}^{\nu}),\label{eq:a2f}
\end{equation}
where $N(E_{F})$ is the density of states at $E_{F}$, $\omega_{\mathbf{q}}^{\nu}$ is phonon frequency, $\epsilon_{\mathbf{k}}^{n}$
is electronic energy, and $g_{\mathbf{k},\mathbf{k}+\mathbf{q}}^{\nu,n,m}$
is electron-phonon coupling matrix element. The total electron-phonon
coupling strength  is then 
\begin{equation}
\lambda=2\int_{0}^{\infty}\frac{\alpha^{2}F(\omega)}{\omega}d\omega.\label{eq:lambda}
\end{equation}
The calculated $\lambda$  for the two samples are 0.81 and 0.96, respectively. We estimated $T_{c}$ based on the modified McMillan  formula \cite{Allen-Dynes}:
 \begin{equation}
T_{C}=\frac{\omega_{log}}{1.2}\exp\left(-\frac{1.04(1+\lambda)}{\lambda-\mu^{*}-0.62\lambda\mu^{*}}\right)\text{,}\label{eq:TC}
\end{equation}
 where  the Coulomb pseudopotential $\mu^{*}$ is set to a typical value of $\mu^{*}=0.1$.   The logarithmically averaged characteristic phonon frequency $\omega_{log}$
is defined as
\begin{equation}
\omega_{log}=\exp\left(\frac{2}{\lambda}\int\frac{d\omega}{\omega}\alpha^{2}F(\omega)\log\omega\right).
\end{equation}
The calculated $\omega_{log}$ for the two samples are 136.7 and 98.4 K, respectively. Using those parameters, we can estimate that the superconductivity with $T_{c}$ of $6\sim7$ K can emerge in the hole doped $1T$-TaS$_{2}$. To observe the superconductivity we predicted, the doping induced metallic network between CCDW domain in NCCDW or other metastable phase must be percolated, which requires experimental devices with ability to dope more holes. Recently, Suda et al. suggested a high level hole doping technique using photoactive electric double layer \cite{Suda-lightEDL}, which might be applied to verify our prediction.

\section{Conclusion}
In conclusion, based on the first-principles calculation, we simulated the carrier doping effect in the bulk and monolayer $1T$-TaS$_{2}$. We found that CCDW is stable upon electron doping, while hole doping can notably suppress the CCDW. According to our analysis, the different mechanisms of the reported electric and photoelectric manipulations of CCDW in $1T$-TaS$_{2}$ are figured out: Light  or positive perpendicular electric field  induced hole doping significantly increases the energy of CCDW, so that the system transforms to NCCDW or similar metastable state. Although the CCDW distortion is more stable upon the in-plane electric field induced electron injection, some accompanied effects can drive the system to cross over the energy barrier from CCDW to  NCCDW or similar metastable state.  We also estimated that a potential superconductivity with $T_{c}$ of $6\sim7$ K can be introduced by hole doping in the system. By tuning the carrier density, controllable switching of different states such as CCDW/Mott insulating state, metallic state, and even the superconducting state can be realized in $1T$-TaS$_{2}$, which makes the novel material have a very promising application in the future electronic devices. 

\begin{acknowledgments}
We thank Prof. Z. G. Sheng and Dr. Y. Z. Zhang for helpful discussion. This work was supported by the National Key Research and Development Program under Contract No. 2016YFA0300404, the National Nature Science Foundation of China (Grant No. 11674326, 11304320, 11274311, 11404340, 1408085MA11, and U1232139).
\end{acknowledgments}

\end{document}